\documentclass{article}
\def\Ref#1{(\ref{#1})}
\usepackage{amsmath}
\usepackage{cite}
\newcommand{\rd}{\mathrm{d}}

\newcommand{\dy}{\mathrm{dy}}

\begin{document}
\begin{titlepage}
\begin{center}
{\large\textbf{Nonuniform autonomous one-dimensional exclusion
nearest neighbor reaction diffusion models, II}}

\vspace*{2\baselineskip}
{\sffamily Amir Aghamohammadi\footnote
{mohamadi@alzahra.ac.ir} \& Mohammad Khorrami
\footnote{mamwad@mailaps.org}}

\vspace*{2\baselineskip}
\textit{Department of Physics, Alzahra University, Tehran 19384, IRAN}
\end{center}

\vspace*{2\baselineskip}


\begin{abstract}\noindent
In a recent article the most general non-uniform reaction-diffusion
models on a one-dimensional lattice with boundaries were considered,
for which the time evolution equations of correlation functions are
closed and the stationary profile can be obtained using a
transfer-matrix method. Here models are investigated for which
also the equation of relaxation towards the stationary profile could
be solved through a similar transfer-matrix method. A classification
is given, and dynamical phase transitions are studied.

\end{abstract}
\end{titlepage}
\newpage
\section{Introduction}
Most studies on reaction-diffusion models are focused on
uniform lattices where reaction rates are site-independent. Among
the simplest generalizations beyond a uniform system is a lattice
with alternating rates. In \cite{SchSch02,SchSch03,MObZ},
the steady state configurational probabilities of an Ising spin chain driven
out of equilibrium by a coupling to two heat baths has been investigated.
An example is a one-dimensional Ising model on a ring, in which the evolution
is according to a generalization of Glauber rates, such that spins at
even (odd) lattice sites experience a temperature $T_\mathrm{e}$
($T_\mathrm{o}$). In \cite{MSZ} the dynamics of a reaction-diffusion model
with spatially alternating rates has been discussed.
The response function to an infinitesimal magnetic field
for the Ising–Glauber model with arbitrary exchange couplings has been
studied in \cite{Chatelain}. In \cite{GO}, relaxation in the kinetic Ising
model on an alternating isotopic chain has been discussed. The critical
dynamics of a disordered Ising ferromagnetic chain with two
coupling constants for Glauber dynamics was studied in \cite{DKMS}.

In \cite{ihg}, a non-uniform extension of the Glauber model on a
one-dimensional lattice with boundaries was investigated. In this model
the reaction rates are based on detailed balance. There
the expectation values of spins in an Ising model with
nonuniform coupling constants were studied. A transfer matrix method
was used to study the steady state behavior of the system in
the thermodynamic limit. Different (static) phases of this system were
studied, and a closed form was obtained for this transfer matrix.
In \cite{nu}, the dynamics and possible dynamical phase transitions of
the non-uniform extension of the Glauber model were investigated. It was
shown that there are two phases: the fast phase and the slow phase.
In the fast phase the relaxation time is independent of the reaction rates
at the boundaries. Changing continuously the reaction rates, the system
may experience a phase transition. In the slow phase
the relaxation time changes with reaction rates at boundaries.

In \cite{dhg} the most general non-uniform exclusion nearest neighbor
reaction-diffusion models on a one-dimensional lattice with boundaries,
were considered, for which the evolution equations of the
one-point functions are closed and the stationary profile of the
system can be obtained using a transfer-matrix method.
It was shown that the criterion that the evolution equations be
closed is the same as the case of uniform models. Such models are
called autonomous. To be able to obtain a closed form for the
transfer matrix, however, additional conditions are to be satisfied.
Models satisfying these additional conditions as well, are called
superautonomous.

In this article we want to study the dynamics and possible
dynamical phase transitions of superautonomous models, using
a transfer-matrix method. The scheme of the paper is as follows.
In section 2, a brief review of the formalism is given. In section 3,
the evolution equations governing the expectation values of
the number operators are studied. It is seen that solvability through
the transfer-matrix method requires additional criteria, apart from
those of the superautonomy conditions obtained in \cite{dhg}.
In section 5 a classification is introduced for superautonomous
models for which the relaxation towards the stationary profile of
the of expectation values of the number operators is solvable through
the transfer-matrix method. In section 5, some examples are explicitly
discussed. Section 6 is devoted to the concluding remarks.

\section{Exclusion nearest-neighbor reaction-diffusion models with nonuniform reaction rates}
Consider a one-dimensional lattice with $(L+1)$ sites, numbered
from $0$ to $L$. Each site is either empty or occupied with one
particle. The evolution of the system is governed by a Hamiltonian
${\mathcal H}$ of the form,
\begin{equation}\label{dnu.01}
{\mathcal H} ={\mathcal H}'_0+\left(\sum_\alpha{\mathcal
H}_\alpha\right)+{\mathcal H}'_L,
\end{equation}
where ${\mathcal H}_\alpha$ corresponds to the link $\alpha$:
\begin{equation}\label{dnu.02}
{\mathcal H}_\alpha=\underbrace{1\otimes\cdots\otimes\cdots
1}_{\alpha-\mu}\otimes\,
H_\alpha\otimes\underbrace{1\otimes\cdots\otimes\cdots
1}_{L-\alpha-\mu},
\end{equation}
and
\begin{equation}\label{dnu.03}
\mu:=\frac{1}{2}.
\end{equation}
The link $\alpha$ links the sites $(\alpha-\mu)$ and
$(\alpha+\mu)$, so that $\alpha\pm\mu$ are integers. $H_\alpha$ is
a linear operator acting on a four dimensional space (the
configuration space corresponding to the sites $(\alpha-\mu)$ and
$(\alpha+\mu)$) with a basis
$\{e_{0\,0},e_{0\,1},e_{1\,0},e_{1\,1}\}$, where $0$ denotes
vacancy and $1$ denotes occupied site. Also,
\begin{align}\label{dnu.04}
{\mathcal H}'_0&=
H'_0\otimes\underbrace{1\otimes\cdots\otimes\cdots 1}_L, \nonumber\\
{\mathcal H}'_L&=\underbrace{1\otimes\cdots\otimes\cdots 1}_L
\otimes H'_L,
\end{align}
where $H'_0$ and $H'_L$ are linear operators acting on two
dimensional spaces (the configuration spaces corresponding to the
sites $0$ and $L$, respectively) with bases $\{e_0,e_1\}$.

The number operator in the site $i$ is denoted by $n_i$:
\begin{equation}\label{dnu.05}
n_i=\underbrace{1\otimes\cdots\otimes\cdots 1}_i \otimes\, n
\otimes\underbrace{1\otimes\cdots\otimes\cdots 1}_{L-i}.
\end{equation}
$n$ is an operator acting on a two dimensional space with the
basis $\{e_0,e_1\}$. The matrix form of $n$ in this basis is
\begin{equation}\label{dnu.06}
n^a{}_b=\delta^a_1\,\delta^1_b.
\end{equation}

The evolution equation for the expectation value of an observable
$Q$ is
\begin{equation}\label{dnu.07}
\frac{\rd}{\rd t}\langle Q\rangle=\langle Q\,{\mathcal H}\rangle,
\end{equation}
where
\begin{equation}\label{dnu.08}
\langle Q\rangle=S\,Q\,\Psi,
\end{equation}
$\Psi$ is the ($2^{L+1}$ dimensional) probability vector
describing the system and $S$ is the covector
\begin{equation}\label{dnu.09}
S:=\underbrace{s\otimes\cdots\otimes\,s}_L,
\end{equation}
and
\begin{equation}\label{dnu.10}
s_a=1.
\end{equation}

The system is called autonomous, iff the Hamiltonian is so that
the evolution of the expectation values of $n_i$ is closed in
terms of the expectation values of $n_j$'s. In the evolution
equation for the expectation value of $n_i$, the expectation
values of $n_{i-1}$, $n_i$, $n_{i+1}$, $(n_{i-1}\,n_i)$, and
$(n_i\,n_{i+1})$ occur. It is seen that the criterion that the
coefficients of the last two vanish, is
\begin{align}\label{dnu.11}
s_a\,[H_{i-\mu}\,r\otimes\,r]^{a\,1}=&
0,\nonumber\\
s_a\,[H_{i+\mu}\,r\otimes\,r]^{1\,a}=& 0,
\end{align}
respectively, where
\begin{equation}\label{dnu.12}
r^a=-\delta^a_0+\delta^a_1.
\end{equation}
\Ref{dnu.11} should hold for all $i$'s, in order that the system be
autonomous. So one can rewrite it like
\begin{align}\label{dnu.13}
s_a\,[H_{\alpha}\,r\otimes\,r]^{a\,1}&=0,\nonumber\\
s_a\,[H_{\alpha}\,r\otimes\,r]^{1\,a}&= 0.
\end{align}
It is seen that this condition is the same as the corresponding
condition for uniform lattices, written for each link separately.
Provided that this condition holds, one arrives at
\begin{equation}\label{dnu.14}
\!\frac{\rd}{\rd t}\langle n_i\rangle=\eta_{i-\mu}\,\langle
n_{i-1}\rangle+\theta_{i+\mu}\,\langle n_{i+1}\rangle
+(\kappa_{i-\mu}+\nu_{i+\mu})\,\langle n_i\rangle
+(\xi_{i-\mu}+\sigma_{i+\mu}), \!\!\!\quad 0<i<L,
\end{equation}
where
\begin{align}\label{dnu.15}
\eta_\alpha&:=s_a\,(H_\alpha)^{a\,1}{}_{b\,0}\,r^b,\nonumber\\
\theta_\alpha&:=s_a\,(H_\alpha)^{1\,a}{}_{0\,b}\,r^b,\nonumber\\
\kappa_\alpha&:=s_a\,(H_\alpha)^{a\,1}{}_{0\,b}\,r^b,\nonumber\\
\nu_\alpha&:=s_a\,(H_\alpha)^{1\,a}{}_{b\,0}\,r^b,\nonumber\\
\xi_\alpha&:=s_a\,(H_\alpha)^{a\,1}{}_{0\,0},\nonumber\\
\sigma_\alpha&:=s_a\,(H_\alpha)^{1\,a}{}_{0\,0}.
\end{align}

For the boundary sites (the sites $0$ and $L$), one has
\begin{align}\label{dnu.16}
\frac{\rd}{\rd t}\langle n_0\rangle&=\theta_\mu\,\langle
n_1\rangle -({H'_0}^0{}_1+{H'_0}^1{}_0-\nu_\mu)\,\langle
n_0\rangle +({H'_0}^1{}_0+\sigma_\mu),\\ \label{dnu.17}
\frac{\rd}{\rd t}\langle n_L\rangle&=\eta_{L-\mu}\,\langle
n_{L-1}\rangle
-(-\kappa_{L-\mu}+{H'_L}^0{}_1+{H'_L}^1{}_0)\,\langle n_L\rangle
+(\xi_{L-\mu}+{H'_L}^1{}_0).
\end{align}

\section{Relaxation towards the static solution}
The evolution equations governing $\langle n_i\rangle_\dy$ (the
difference of $\langle n_i\rangle$ with its static value) are the
homogeneous parts of \Ref{dnu.14}, \Ref{dnu.16}, and \Ref{dnu.17},
which can be written like
\begin{equation}\label{dnu.18}
\frac{\rd}{\rd t}\langle n_j\rangle_\dy=h_j^l\,\langle
n_l\rangle_\dy.
\end{equation}
The eigenvalues and eigenvectors of the operator $h$ satisfy
\begin{align}\label{dnu.19}
E\,x_j&=\eta_{j-\mu}\,x_{j-1}+\theta_{j+\mu}\,x_{j+1}
+(\kappa_{j-\mu}+\nu_{j+\mu})\,x_j, \quad 0<j<L,\\
\label{dnu.20} E\,x_0&=\theta_\mu\,x_1
-({H'_0}^0{}_1+{H'_0}^1{}_0-\nu_\mu)\,x_0,\\
\label{dnu.21} E\,x_L&=\eta_{L-\mu}\,x_{L-1}
-(-\kappa_{L-\mu}+{H'_L}^0{}_1+{H'_L}^1{}_0)\,x_L,
\end{align}
where $E$ is the eigenvalue and $x$ is the eigenvector.
The solution to \Ref{dnu.19} is
\begin{equation}\label{dnu.22}
X_{j+\mu}=\tilde D_j\,X_{j-\mu},
\end{equation}
where
\begin{equation}\label{dnu.23}
X_\alpha:=\begin{bmatrix}x_{\alpha-\mu}\\
x_{\alpha+\mu}\end{bmatrix},
\end{equation}
and
\begin{equation}\label{dnu.24}
\tilde D_j:=\begin{bmatrix}0&1 \\ & \\
\displaystyle{-\frac{\eta_{j-\mu}}{\theta_{j+\mu}}}&
\displaystyle{-\frac{\kappa_{j-\mu}+\nu_{j+\mu}-E}{\theta_{j+\mu}}}\end{bmatrix}.
\end{equation}
Using these, one arrives at
\begin{equation}\label{dnu.25}
X_\alpha=\tilde D_{\alpha\,\beta}\,X_\beta,
\end{equation}
where
\begin{equation}\label{dnu.26}
\tilde D_{\alpha\,\beta}:=\tilde D_{\alpha-\mu}\cdots\,\tilde
D_{\beta+\mu}.
\end{equation}
\Ref{dnu.20} and \Ref{dnu.21} lead to
\begin{equation}\label{dnu.27}
X_\mu\propto \begin{bmatrix}\theta_\mu\\ \\
E-\kappa_\mu-\nu_\mu+\zeta\end{bmatrix},
\end{equation}
and
\begin{equation}\label{dnu.28}
X_{L-\mu}\propto \begin{bmatrix}E-\kappa_{L-\mu}-\nu_{L-\mu}+
\lambda\\
\\\eta_{L-\mu}\end{bmatrix},
\end{equation}
respectively, where
\begin{align}\label{dnu.29}
\zeta&:={H'_0}^1{}_0+{H'_0}^0{}_1+\kappa_\mu,\nonumber\\
\lambda&:={H'_L}^1{}_0+{H'_L}^0{}_1+\nu_{L-\mu}.
\end{align}
Defining
\begin{equation}\label{dnu.30}
V:=\begin{bmatrix}\theta_\mu\\ \\
E-\kappa_\mu-\nu_\mu+\zeta\end{bmatrix},
\end{equation}
and
\begin{equation}\label{dnu.31}
W:=\begin{bmatrix}-\eta_{L-\mu}&
E-\kappa_{L-\mu}-\nu_{L-\mu}+\lambda\end{bmatrix},
\end{equation}
it is seen that the eigenvalues can be obtained through
\begin{equation}\label{dnu.32}
W\,\tilde D_{L-\mu\,\mu}\,V=0.
\end{equation}

Things become simpler if one can write $\tilde D_j$ as
\begin{equation}\label{dnu.33}
\tilde
D_j:=\tilde\Sigma_{j+\mu}\,\tilde\Delta_j\,\tilde\Sigma^{-1}_{j-\mu},
\end{equation}
where $\tilde\Delta_i$ is diagonal, and $\tilde\Sigma_\alpha$
depends on only the parameters corresponding to the link $\alpha$.
Putting
\begin{equation}\label{dnu.34}
\tilde\Sigma_\alpha=\begin{bmatrix}\tilde a_\alpha &\tilde b_\alpha \\
\tilde c_\alpha &\tilde d_\alpha\end{bmatrix},
\end{equation}
and
\begin{equation}\label{dnu.35}
\tilde \Delta_j=\begin{bmatrix}
\tilde A_j& 0 \\ & \\
0 & \tilde B_j\end{bmatrix},
\end{equation}
in \Ref{dnu.33}, exactly similar to \cite{nu} one arrives at
\begin{align}\label{dnu.36}
\tilde A_j&=\frac{\tilde c_{j-\mu}}{\tilde a_{j+\mu}},\\
\label{dnu.37}\tilde B_j&=\frac{\tilde d_{j-\mu}}{\tilde
b_{j+\mu}},
\end{align}
and
\begin{align}\label{dnu.38}
\tilde a_\alpha\,\tilde b_\alpha&=
\tilde\phi\,\theta_\alpha\,\tilde\varsigma_\alpha,\\
\label{dnu.39} \tilde c_\alpha\,\tilde d_\alpha&=
\tilde\phi\,\eta_\alpha\,\tilde\varsigma_\alpha,\\
\label{dnu.40} \tilde a_\alpha\,\tilde d_\alpha&=
(-\tilde\phi\,\tilde\kappa_\alpha+\tilde\psi)\,\tilde\varsigma_\alpha,\\
\label{dnu.41} \tilde b_\alpha\,\tilde c_\alpha&=
(-\tilde\phi\,\tilde\nu_\alpha-\tilde\psi)\,\tilde\varsigma_\alpha,
\end{align}
where
\begin{equation}\label{dnu.42}
\tilde\varsigma_\alpha:=\tilde a_\alpha\,\tilde d_\alpha-\tilde
b_\alpha\,\tilde c_\alpha,
\end{equation}
and $\tilde\phi$ and $\tilde\psi$ are link independent. The
parameters $\tilde\kappa$ and $\tilde\nu$ are related to $\kappa$
and $\nu$ through
\begin{equation}\label{dnu.43}
\tilde\kappa_{j-\mu}+\tilde\nu_{j+\mu}:=
\kappa_{j-\mu}+\nu_{j+\mu}-E.
\end{equation}
The consistency of \Ref{dnu.38} to \Ref{dnu.41} requires that
$(\tilde\kappa-\tilde\nu)$ and
$[(\tilde\kappa+\tilde\nu)^2-4\,\eta\,\theta]$ be constant (link
independent). Using
\begin{equation}\label{dnu.44}
(\tilde\kappa_\alpha+\tilde\nu_\alpha)^2-
4\,\eta_\alpha\,\theta_\alpha=(\kappa_\alpha+\nu_\alpha)^2-
4\,\eta_\alpha\,\theta_\alpha-2\,E\,(\kappa_\alpha+\nu_\alpha)+E^2,
\end{equation}
it turns out that in order that the left hand side be link
independent for all of the values of $E$, both $(\kappa+\nu)$ and
$(\eta\,\theta)$ should be link independent. Combining these with
the link independence of $(\kappa-\nu)$, it is seen that the
necessary and sufficient condition for the possibility of the
decomposition \Ref{dnu.33} is that $\kappa$, $\nu$, and
$(\eta\,\theta)$ be link independent. This is one condition more
than the conditions for superautonomy (\cite{nu}), where it was
necessary and sufficient that $(\kappa-\nu)$ and
$(\kappa\,\nu-\eta\,\theta)$ be link independent.

Assuming that \Ref{dnu.33} holds, one arrives at
\begin{equation}\label{dnu.45}
\tilde D_{L-\mu\,\mu}=\begin{bmatrix}\frac{\tilde c_\mu\,\tilde
d_\mu}{\tilde\varsigma_\mu}\,
(\Xi_{L-3\,\mu\;3\,\mu}-\Upsilon_{L-3\,\mu\;3\,\mu}) &
\frac{\tilde a_\mu\,\tilde b_\mu}{\tilde\varsigma_\mu}\,
(\Upsilon_{L-3\,\mu\;\mu}-\Xi_{L-3\,\mu\;\mu})\\ & \\
\frac{\tilde c_\mu\,\tilde d_\mu}{\tilde\varsigma_\mu}\,
(\Xi_{L-\mu\;3\,\mu}-\Upsilon_{L-\mu\;3\,\mu}) & \frac{\tilde
a_\mu\,\tilde b_\mu}{\tilde\varsigma_\mu}\,
(\Upsilon_{L-\mu\;\mu}-\Xi_{L-\mu\;\mu}),
\end{bmatrix}
\end{equation}
where
\begin{align}\label{dnu.46}
\Xi_{\alpha\,\beta}&:=\frac{\tilde c_\alpha}{\tilde
a_\alpha}\cdots\frac{\tilde c_\beta}{\tilde a_\beta},\nonumber\\
\Upsilon_{\alpha\,\beta}&:=\frac{\tilde d_\alpha}{\tilde
b_\alpha}\cdots\frac{\tilde d_\beta}{\tilde b_\beta}.
\end{align}
Defining $Z_1$ through
\begin{equation}\label{dnu.47}
\frac{\tilde c_\alpha}{\tilde a_\alpha}=:Z_1\,\frac{\eta_\alpha}{
\sqrt{|\eta_\alpha\,\theta_\alpha|}},
\end{equation}
it is seen that $Z_1$ is link independent and
\begin{equation}\label{dnu.48}
\frac{\tilde d_\alpha}{\tilde b_\alpha}=:Z_2\,\frac{\eta_\alpha}{
\sqrt{|\eta_\alpha\,\theta_\alpha|}},
\end{equation}
where
\begin{equation}\label{dnu.49}
Z_1\,Z_2=\mathrm{sgn}(\eta_\alpha\,\theta_\alpha).
\end{equation}
One then arrives at
\begin{equation}\label{dnu.50}
E=\kappa+\nu+\sqrt{|\eta\,\theta|}\,(Z_1^{-1}+Z_2^{-1}),
\end{equation}
where use has been made of the fact that $\kappa$, $\nu$, and
$(\eta\,\theta)$ are link-independent. It is seen that
\Ref{dnu.32} is equivalent to
\begin{equation}\label{dnu.51}
W\,\begin{bmatrix}\displaystyle{\frac{1}{\eta_{L-\mu}\,\theta_\mu}\,
(Z_1^{L-2}-Z_2^{L-2})} &
\displaystyle{-\frac{1}{\eta_{L-\mu}\,\sqrt{|\eta\,\theta|}}\,
(Z_1^{L-1}-Z_2^{L-1})}\\ & \\
\displaystyle{\frac{1}{\theta_\mu\,\sqrt{|\eta\,\theta|}}\,
(Z_1^{L-1}-Z_2^{L-1})} &\displaystyle{-\frac{1}{|\eta\,\theta|}\,
(Z_1^L-Z_2^L)}\end{bmatrix}\,V=0,
\end{equation}
or
\begin{align}\label{dnu.52}
0&=Z_2^L\,\left(\frac{1}{Z_1}+\frac{\lambda}{\sqrt{|\eta\,\theta|}}\right)
\,\left(\frac{1}{Z_1}+\frac{\zeta}{\sqrt{|\eta\,\theta|}}\right)\nonumber\\
&-
Z_1^L\,\left(\frac{1}{Z_2}+\frac{\lambda}{\sqrt{|\eta\,\theta|}}\right)
\,\left(\frac{1}{Z_2}+\frac{\zeta}{\sqrt{|\eta\,\theta|}}\right).
\end{align}
In the thermodynamic limit ($L\to\infty$), all of the unimodular values
of $Z_2$ are solutions to \Ref{dnu.52}. If there are no further solutions,
the relaxation time would be $\tau_\mathrm{fast}$:
\begin{equation}\label{dnu.53}
\tau_\mathrm{fast}=[-\kappa-\nu-2\,\mathrm{Re}(\sqrt{\eta\,\theta})]^{-1}.
\end{equation}
Things change if there are nonunimodular solutions for $Z_2$ as well.
In that case, one can assume (without loss of generality) that $|Z_2|$
is larger than one. Then \Ref{dnu.52} in the thermodynamic limit leads to
\begin{equation}\label{dnu.54}
\left(\frac{1}{Z_1}+\frac{\lambda}{\sqrt{|\eta\,\theta|}}\right)
\,\left(\frac{1}{Z_1}+\frac{\zeta}{\sqrt{|\eta\,\theta|}}\right)=0,
\end{equation}
or
\begin{align}\label{dnu.55}
Z_1^{-1}&=-\frac{\zeta}{\sqrt{|\eta\,\theta|}},\\ \label{dnu.56}
Z_1^{-1}&=-\frac{\lambda}{\sqrt{|\eta\,\theta|}}.
\end{align}
Each of \Ref{dnu.55} or \Ref{dnu.56} are acceptable, of course,
provided the absolute value of the right-hand side is
larger than one. Putting these in \Ref{dnu.50}, one
arrives at the following time scales $[-\mathrm{Re}(E)]^{-1}$:
\begin{align}\label{dnu.57}
\tau_\zeta&=[-\kappa-\nu+\zeta+\zeta^{-1}\,\eta\,\theta]^{-1},\nonumber\\
\tau_\lambda&=[-\kappa-\nu+\lambda+\lambda^{-1}\,\eta\,\theta]^{-1}.
\end{align}
One notes that each of $\tau_\zeta$ and $\tau_\lambda$ are larger than
$\tau_\mathrm{fast}$, if the right-hand sides of \Ref{dnu.55} and \Ref{dnu.56}
are larger than one, respectively. So one arrives at the following expression
for the relaxation time ($\tau$).
\begin{equation}\label{dnu.58}
\tau=\begin{cases}
\tau_\mathrm{fast},& (-\lambda<\sqrt{|\eta\,\theta|})\wedge(-\zeta<\sqrt{|\eta\,\theta|})\\
\tau_\zeta,& (-\lambda<\sqrt{|\eta\,\theta|})\wedge(-\zeta>\sqrt{|\eta\,\theta|})\\
\tau_\lambda,& (-\lambda>\sqrt{|\eta\,\theta|})\wedge(-\zeta<\sqrt{|\eta\,\theta|})\\
\mathrm{max}(\tau_\lambda,\tau_\zeta),&
(-\lambda>\sqrt{|\eta\,\theta|})\wedge(-\zeta>\sqrt{|\eta\,\theta|})
\end{cases}.
\end{equation}
The first case corresponds to the fast phase, where the relaxation time
does not depend on the boundary rates. The other cases correspond to the slow
phase, where the relaxation time does depend on the boundary rates.

\section{Classification of the solvable models}
The local Hamiltonian has 12 independent parameters. Of these
12 parameters, only 8 enter the evolution equations of
the expectation values of the number operators. The criteria
for the autonomy of the system are the two equations \Ref{dnu.13}.
The system is called superautonomous, if there is a basis corresponding
to each link, which diagonalizes all one point transfer-matrices
corresponding to the equation for the time independent configuration
of one point functions (hence diagonalizing the whole transfer matrix).
That condition could be written as the existence of matrices $\Delta$ and
$\Sigma$ such that
\begin{equation}\label{dnu.59}
D_i=\Sigma_{i+\mu}\,\Delta_i\,\Sigma_{i-\mu},
\end{equation}
as discussed in \cite{dhg}.
The additional criteria for the system to be superautonomous and
its relaxation be solvable through the transfer-matrix method are
that $\kappa$, $\nu$, and $(\eta\,\theta)$ be link-independent.
All of these 5 conditions can be written in terms of the following
8 nonnegative independent combinations of the rates.
\begin{align}\label{dnu.60}
A_\alpha&:=s_a\,(H_\alpha)^{a\,1}{}_{0\,0},\nonumber\\
B_\alpha&:=s_a\,(H_\alpha)^{1\,a}{}_{0\,0},\nonumber\\
D_\alpha&:=s_a\,(H_\alpha)^{a\,0}{}_{0\,1},\nonumber\\
F_\alpha&:=s_a\,(H_\alpha)^{1\,a}{}_{0\,1},\nonumber\\
G_\alpha&:=s_a\,(H_\alpha)^{0\,a}{}_{1\,0},\nonumber\\
I_\alpha&:=s_a\,(H_\alpha)^{a\,1}{}_{1\,0},\nonumber\\
K_\alpha&:=s_a\,(H_\alpha)^{0\,a}{}_{1\,1},\nonumber\\
L_\alpha&:=s_a\,(H_\alpha)^{a\,0}{}_{1\,1}.
\end{align}
The model is autonomous iff
\begin{align}\label{dnu.61}
A_\alpha+D_\alpha&=I_\alpha+L_\alpha,\nonumber\\
B_\alpha+G_\alpha&=F_\alpha+K_\alpha.
\end{align}
One has
\begin{align}\label{dnu.62}
\eta_\alpha&=I_\alpha- A_\alpha,\nonumber\\
\kappa&=-A_\alpha-D_\alpha,\nonumber\\
\nu&=-B_\alpha-G_\alpha,\nonumber\\
\theta_\alpha&=F_\alpha-B_\alpha.
\end{align}
Fixing $\kappa$, $\eta_\alpha$, $\nu$, and $\theta_\alpha$, using
\Ref{dnu.61} and \Ref{dnu.62}, one is still left with two degrees of freedom.
It should be noted, however, that $(\eta_\alpha\,\theta_\alpha)$ should be
link-independent, and $(\eta_\alpha-\kappa)$, and $(\theta_\alpha-\nu)$ are
nonnegative. Then the parameters entering evolution equations are
$\kappa$, $\nu$, $\eta_\alpha$, $\theta_\alpha$, $A_\alpha$, and $B_\alpha$.
The local Hamiltonian, however, still contains two more free parameters
which do not enter the time evolution equations of the expectation values of
the number operators.

\section{Examples}
Consider a lattice each site of which is either empty ($\circ$) or
full ($\bullet$). Two examples are studied here.
\subsection{The diffused voting model}
The reactions on a link are
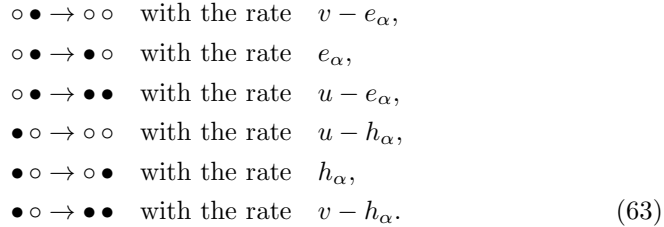
\begin{align}\label{dnu.63}
\circ\,\bullet\to\circ\,\circ &\mbox{\quad with the rate\quad} v-e_\alpha,\nonumber\\
\circ\,\bullet\to\bullet\,\circ &\mbox{\quad with the rate\quad} e_\alpha,\nonumber\\
\circ\,\bullet\to\bullet\,\bullet &\mbox{\quad with the rate\quad} u-e_\alpha,\nonumber\\
\bullet\,\circ\to\circ\,\circ &\mbox{\quad with the rate\quad} u-h_\alpha,\nonumber\\
\bullet\,\circ\to\circ\,\bullet &\mbox{\quad with the rate\quad} h_\alpha,\nonumber\\
\bullet\,\circ\to\bullet\,\bullet &\mbox{\quad with the rate\quad} v-h_\alpha.
\end{align}
The rates of injection and extraction of particles in the first (final) sites
are $a$ and $b$ ($a'$ and $b'$), respectively. The rates $u$ and $v$ are
link-independent, but the diffusion rates, $h_\alpha$ and $e_\alpha$,
may be link-dependent. Using \Ref{dnu.15} and \Ref{dnu.29} one arrives at
\begin{align}\label{dnu.64}
\eta&=v,\nonumber\\
\theta&=u,\nonumber\\
\kappa&=-v,\nonumber\\
\nu&=-u,\nonumber\\
\xi&=0,\nonumber\\
\sigma&=0,\nonumber\\
\zeta&=a+b-v,\nonumber\\
\lambda&=a'+b'-u.
\end{align}
Although the diffusion rates are link-dependent, all the parameters
entering the time evolution equation for $\langle n_i\rangle_\dy$ are
link-independent. The time evolution equations are exactly the same as
those obtained for the corresponding uniform model \cite{29}. There
using a different method, it was shown that the dynamical
phase transition is controlled by the reaction rates at the boundaries.

Using \Ref{dnu.53}, \Ref{dnu.57}, and \Ref{dnu.64}, one has
\begin{align}\label{dnu.65}
\tau_\mathrm{fast}&=\frac{1}{(\sqrt{u}+\sqrt{v})^2},\nonumber\\
\tau_\zeta&=\left(a+b+u+\frac{u\,v}{a+b-v}\right)^{-1},\nonumber\\
\tau_\lambda&=\left(a'+b'+v+\frac{u\,v}{a'+b'-u}\right)^{-1}.
\end{align}
It is seen from \Ref{dnu.58} that
\begin{equation}\label{dnu.66}
\tau=\begin{cases}\tau_\zeta,&(a+b)<(v-\sqrt{u\,v})\\
\tau_\lambda,&(a'+b')<(u-\sqrt{u\,v})\\
\tau_\mathrm{fast},&\mathrm{otherwise}
\end{cases}.
\end{equation}
The system is in the fast phase (the last case), when
the boundary rates are high enough, so that it is the bulk reaction rates
that determine the relaxation, and goes to the slow phase when
the boundary reaction rates are less than some critical value.
\subsection{Diffusion and annihilation}
The reactions on a link are
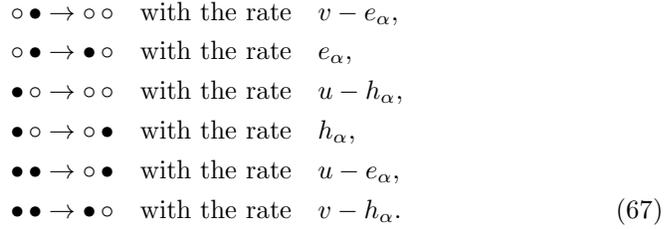
\begin{align}\label{dnu.67}
\circ\,\bullet\to\circ\,\circ &\mbox{\quad with the rate\quad} v-e_\alpha,\nonumber\\
\circ\,\bullet\to\bullet\,\circ &\mbox{\quad with the rate\quad} e_\alpha,\nonumber\\
\bullet\,\circ\to\circ\,\circ &\mbox{\quad with the rate\quad} u-h_\alpha,\nonumber\\
\bullet\,\circ\to\circ\,\bullet &\mbox{\quad with the rate\quad} h_\alpha,\nonumber\\
\bullet\,\bullet\to\circ\,\bullet &\mbox{\quad with the rate\quad} u-e_\alpha, \nonumber \\
\bullet\,\bullet\to\bullet\,\circ &\mbox{\quad with the rate\quad} v-h_\alpha.
\end{align}
The rates of injection and extraction of particles in the first (final) sites
are $a$ and $b$ ($a'$ and $b'$), respectively. The rates $u$ and $v$ are
link-independent, but $h_\alpha$ and $e_\alpha$ could be link-dependent, with
their product link-independent:
\begin{equation}\label{dnu.68}
e_\alpha\,h_\alpha=m.
\end{equation}
Also each of $e_\alpha$ and $h_\alpha$ should be less than or equal to
each of $u$ and $v$, resulting in
\begin{equation}\label{dnu.69}
m\leq u\,v.
\end{equation}
Using \Ref{dnu.15} and \Ref{dnu.29} one arrives at
\begin{align}\label{dnu.70}
\eta_\alpha&=h_\alpha,\nonumber\\
\theta_\alpha&=e_\alpha,\nonumber\\
\kappa&=-v,\nonumber\\
\nu&=-u,\nonumber\\
\xi&=0,\nonumber\\
\sigma&=0,\nonumber\\
\zeta&=a+b-v,\nonumber\\
\lambda&=a'+b'-u.
\end{align}
Using \Ref{dnu.53}, \Ref{dnu.57}, and \Ref{dnu.70}, one has
\begin{align}\label{dnu.71}
\tau_\mathrm{fast}&=(u+v-2\,\sqrt{m})^{-1},\nonumber\\
\tau_\zeta&=\left(a+b+u+\frac{m}{a+b-v}\right)^{-1},\nonumber\\
\tau_\lambda&=\left(a'+b'+v+\frac{m}{a'+b'-u}\right)^{-1},
\end{align}
and from \Ref{dnu.58}
\begin{equation}\label{dnu.72}
\tau=\begin{cases}
\tau_\mathrm{fast},& (a'+b'>u-\sqrt{m})\wedge(a+b>v-\sqrt{m})\\
\tau_\zeta,& (a'+b'>u-\sqrt{m})\wedge(a+b<v-\sqrt{m})\\
\tau_\lambda,& (a'+b'<u-\sqrt{m})\wedge(a+b>v-\sqrt{m})\\
\mathrm{max}(\tau_\lambda,\tau_\zeta),&(a'+b'>u-\sqrt{m})\wedge(a+b>v-\sqrt{m})
\end{cases}.
\end{equation}
It may be that both $(u-\sqrt{m})$ and $(v-\sqrt{m})$ are positive, so unlike
the previous example it is possible that two phase transitions occur.
\section{Concluding remarks}
Autonomous models are those for which the evolution equations for
the expectation values of the number operators are closed. Recently
the most general autonomous exclusion models with nearest-neighbor
interactions and non-uniform reaction rates on a one-dimensional lattice
were studied. In \cite{dhg}, using a transfer-matrix method, possible
static phase transitions of such systems had been investigated.
Here a similar transfer-matrix method was used to study the dynamics
and possible dynamical phase transitions of superautonomous models.
It was seen that superautonomy does not guarantee that the relaxation of
the system towards its stationary state be solvable using the
transfer-matrix method. One further constraint is needed, which
was found. It was also seen that there are two possible phases,
regarding the relaxation. In the fast phase, the relaxation
does not depend on boundary rates. In this phase the relaxation time is
controlled by only the bulk reaction rates. There may be slow phases
as well. In these cases, the relaxation does receive contributions from the
boundary rates. A classification was presented for the models solvable
through the transfer-matrix method, and two examples were explicitly
stduied.
\\[\baselineskip]
\textbf{Acknowledgement}: This work was partially supported by
the research council of the Alzahra University.

\end{document}